\documentstyle[preprint,aps,version2]{revtex}
\begin{document}
  
\begin{title}
Study of Baryon Antibaryon Rapidity Correlation \\
in $e^{+}e^{-}$ Annihilation by Quark Combination Model
\end{title}

\author{Zong-Guo Si$^1$, 
Qu-Bing Xie$^{2,1}$, Qun Wang$^1$}

\begin{instit}
$^1$ Department of Physics, Shandong University

Jinan, Shandong 250100, P.R. China 

$^2$ Center of Theoretical Physics, CCAST(World Lab)

Beijing 100080, P. R. China
\end{instit}

\begin{abstract}
We use Quark Combination Model to study baryon antibaryon 
rapidity correlation
in $e^{+}e^{-}$ annihilation and compare our predictions 
with the available
data. We find that these results and relevant properties 
are all consistent 
with data. So the evidence to rule out Quark Combination 
Model which has
long been cited in literatures does not exist.
\end{abstract}
\pagestyle{plain}
\section{Introduction}
\label{i}
More than half of the measured baryons in $e^+e^-$ annihilation are
directly produced, and even if they are the decay products of primary
baryons, unlike most of mesons, they almost conserve the rapidity and
the direction of their parents. Hence, the properties of the 
baryon production especially the correlation between baryon($B$) and 
its antibaryon($\overline{B}$) can provide a better probe of the 
hadronization mechanism. But even for the
data now available, the popular models face some difficulties. 
For instance, in order to explain the baryon production and the 
$B\overline{B}$ rapidity correlation, the Lund String 
Fragmentation Model~(LM) has to introduce a problematical vacuum 
excitation of diquark-antidiquark pairs and the popcorn scenario, 
so that it brings at least 7 free 
parameters$^{\cite{Mattig,Angelis,sjostrand}}$. Even so,
it cannot reproduce the octet and decuplet baryon multiplicities 
simultaneously$^{\cite{Opal}}$. Recently, OPAL 
collaboration find that among all of the popular models, the observed 
$B\overline{B}$ rapidity correlation can only be described by 
adjusting the 
$ad~hoc$ popcorn parameter to 95\% in LM$^{\cite{Opalbb}}$, 
but it has no prediction power 
for the energy dependence of the popcorn parameter 
and for the probabilities of the $B\overline{B}$, $BM\overline{B}$, 
$BMM\overline{B}$ etc. configurations.

Original Quark Combination Model~(QCM) was first proposed 
by Annisovich and Bjorken et~al$^{\cite{Anni}}$. It was famous for 
its simple picture and its successful prediction for the percentage 
of vector mesons. One of its great merits is that it treats the 
baryon and meson production in an uniform scheme, so
it describes the baryon production naturally.
But TASSO collaboration studied the proton($p$) 
anti-proton($\overline{p}$)
phase space correlation at $\sqrt{s}~\simeq~30~GeV$ in $e^{+}e^{-}$ 
annihilation early in 1983. They found that the prediction of 
Cerny's Monte Carlo Program~(CMCP) which was alleged to be 
based on QCM showed great
discrepancies with data$^{\cite{TASSO}}$. From then on, although 
QCM is superior to the other models in describing the baryon
production, this conclusion that QCM is ruled out by $B\overline{B}$
rapidity correlation is always cited in the later
literatures$^{\cite{Mattig,Angelis}}$,
since it is believed that compared with the inclusive properties,  
the $B\overline{B}$ rapidity correlation can provide a more 
effective criteria 
to discriminate different models$^{\cite{Mattig,Angelis,sjostrand}}$.

In order to investigate whether the contradiction between the TASSO 
data and the prediction of CMCP is really caused by QCM itself, 
several years ago, we analyzed the $B\overline{B}$ phase space 
correlation from the naive QCM scheme and found that there 
should not be such an inconsistency qualitatively$^{\cite{Xie1}}$.
In the meantime, we developed Quark Production Rule~(QPR) 
and Quark Combination 
Rule~(QCR) in the QCM scheme$^{\cite{Xie2}}$ and use them 
to explain a series of phenomena in $e^{+}e^{-}$ annihilation 
successfully by using much less adjustable parameters. These 
phenomena include the multiplicities of various hadrons, the energy 
dependence of $B/M$ ratio, the multiplicity 
distribution, the so-called spin suppression for baryons, 
high multiplicity of singlet baryons and $B\overline{B}$ flavor 
correlations etc$^{\cite{Xie2,qun,chen,wang,Liang}}$.

In this paper, we study the $B\overline{B}$ rapidity correlation 
by QCM in detail, and discuss some related properties. 
In order to compare
the predictions of QCM with data and with those of other models 
more properly, we implant QCM in JETSET Monte Carlo generator 
to describe 
the hadronization. Then we use this modified JETSET generator
to study the $B\overline{B}$ correlation. We find that the 
$p\overline{p}$ phase space correlation at 
$\sqrt{s} \simeq 30~GeV$ is in agreement with data. Thus the
evidence to exclude QCM which has long been cited in literatures 
does not exist. 
In particular, OPAL collaboration recently presented high statistics
measurements for $\Lambda\overline{\Lambda}$
rapidity correlation at 91~GeV and precisely compared their data with
the popular models including LM, Webber Cluster Fragmentation 
Model~(WM) and UCLA model$^{\cite{Opalbb}}$. We also give the 
prediction for $\Lambda\overline{\Lambda}$ rapidity correlation and
find that our predictions agree with the OPAL data quite well. 
We find that the $B\overline{B}$ correlation is insensitive to 
inclusive quantities as was expected. Additionally, we give 
prediction for some $B/M$ ratios, multiplicities of some strange 
baryons, the differential cross sections of the related baryons 
and the local baryon number compensations.

This paper is organized as follows: In section~\ref{ii}, 
we discuss the Monte Carlo implementation of QCM, give a brief 
description of QPR and QCR, 
and list the relevant formula. In section~\ref{iii}, we give the
predictions of $p\overline{p}$ (at
$\sim30~GeV$) and $\Lambda{\bar\Lambda}$ (at
91 GeV) rapidity correlation, the differential cross sections of
the related baryons and other results. Finally, in section~\ref{iv},
we make a conclusive remarks.

\section{Monte Carlo Implementation of QCM}
\label{ii}
In order to compare the available data with the predictions of a
hadronization model appropriately, a complete 
$e^{+}e^{-} \rightarrow h's$  generator is needed.
It should contain the following four phases:

1. $e^+e^-\rightarrow \gamma/Z^0 \rightarrow q\overline{q}$, 
i.e. $e^{+}e^{-}$ pair converts into a primary 
quark pair $q\overline{q}$ via virtual photon or 
$Z^{0}$. This phase is described by the electro-weak theory.

2. Perturbative phase. It describes the radiation of gluons 
off the primary quarks, and the subsequent parton cascade due 
to gluon splitting into quarks and gluons, and the gluon radiation 
of secondary quarks. It is believed that 
perturbative QCD can describe this phase quantatively. We may use 
the Parton Shower(PS) approach, such as PS model in JETSET (or in
HERWIG) and Color Dipole Model etc. to describe this phase. 
Though each of
them has different evolution parameters, all of them are based on 
AP equation, so the results obtained by different PS models should 
be similar. In this paper, we use the PS Model in JETSET.
 
3. hadronization phase. Up to now, this phase can only be described 
by phenomenological models, such as LM, WM, UCLA model and QCM etc.

4. Unstable hadron decays. The descriptions of this phase in JETSET 
or HERWIG are quite similar except some c, b hadrons.

As is usually done, the Monte Carlo implementation of different
hadronization models can be embedded in the same generator provided 
that the corresponding interface is correctly connected. 
For instance, one can use either LM or Independent Fragmentation 
Model to describe the hadronization
phase in JETSET. Another example is that Buchman et al. replaced LM
with UCLA model in JETSET$^{\cite{buchman}}$. They use this modified
program to reproduce the multiplicities of various hadrons quite 
well. In our work, we substitute QCM for LM in JETSET to describe the
hadronization process. For the sake of comparison with 
LM, UCLA and other models,  we keep other three parts of JETSET 7.3 
unchanged and use the default values for all of the parameters 
emerged therein. So we can investigate the impact of different 
hadronization models on $B\overline{B}$ correlation, 
the baryon number compensation, the $B/M$ ratio, multiplicities of 
strange baryons independently. 

In the following, we first recall QPR and QCR for a color 
singlet system, then simply use the Longitudinal Phase Space 
Approximation~(LPSA) to get
the momentum distribution for primary hadrons in its own system. 
Finally we extend this hadronization scheme to multi-parton states, 
and connect it with the perturbative phase in JETSET 7.3.

\subsection  {QPR and QCR for $q\overline{q}$ system}
\label{m1}

As was mentioned in section~\ref{i}, QPR and QCR have successfully 
explained a set of phenomena in $e^{+}e^{-}$ annihilation. 
Here we briefly introduce them and list the relevant equations~(for 
detail, see ref.~\cite{Xie2}).

In a color singlet system formed by $q\overline{q}$, 
$N$ pairs of quarks can be produced by vacuum excitation via strong 
interaction. We assume that $N$ satisfies Poisson Distribution:
\begin{equation} 
\label{fir}
P(<N>,N-1) = {<N>^{N-1} \over (N-1)!} e^{-<N>}
\end{equation} 
where $<N>$ is the average number of those quark pairs. 
According to QPR, $<N>$ is given by
\begin{equation}
\label{fir1}
<N>=
\sqrt{\alpha^{2}+\beta(W-{{M}_{q}}-{{M}_{\overline{q}}}
+2\overline{m})} -\alpha-1, ~~~~~\\
\alpha=\beta\overline{m}-{{1} \over {4}}
\end{equation}
where $W$ is the invariant mass of the system, $\beta$ is a free 
parameter, $\overline{m}$ is the average mass of newborn quarks, 
and $M_{q}$ and $M_{\overline{q}}$ are the masses of endpoint quark 
and anti-quark. Thus we have $N$ pairs of quarks according to
eqs.~(\ref{fir}),~(\ref{fir1}) (containing one primary quark pair). 

When describing how quarks and antiquarks form hadrons, we find that
all kinds of hadronization models satisfy the near correlation in 
rapidity more or less. Since there is no deep
understanding of the significance and the role of this, the near 
rapidity correlation has not been
used sufficiently. In ref.~\cite{Xie1}, we have shown
that the nearest correlation in rapidity is in agreement with the
fundamental requirements of QCD, and determines QCR completely. 
The rule guarantees that the combination of quarks across more 
than two rapidity gaps never emerges and that $N$ quarks and $N$ 
antiquarks are exactly exhausted
without forming baryonium $qq\overline{q}\overline{q}$ 
and other things. Considering that the quarks and antiquarks are 
stochastically arranged in rapidity space, each order can occur 
with the same probability. Then the probability distribution for 
$N$ quarks and $N$ antiquarks to combine into $M$ mesons, $B$ 
baryons and $B$ anti-baryons according to QCR is given by
\begin{equation}
\label{w1}
{X_{MB} =} {{2N(N!)^{2}(M+2B-1)!} \over {(2N)!M!(B!)^{2}}} 3^{M-1}
\delta_{N,M+3B}
\end{equation}
The average numbers of primary mesons $M(N)$ and baryons $B(N)$ are
\begin{equation}
\label{w2}
\left \{
\begin{array}{l}
M(N)=\sum\limits_{M,B}MX_{MB}(N) \\
B(N)=\sum\limits_{M,B}BX_{MB}(N) 
\end{array}
\right.
\end{equation}
Approximately, in the combination for $N\geq 3$, $M(N)$ and 
baryons $B(N)$ can be well parameterized as linear functions 
of quark number $N$,
\begin{equation}
\label{w3}
\left\{
\begin{array}{l}
M(N)=aN+b\\
B(N)={{(1-a)} \over 3}N -{b \over 3}
\end{array}
\right.
\end{equation}
where $a=0.66$ and $b=0.56$. But for $N<3$, one has
\begin{equation}
\label{w4}
M(N)=N,~~~~B(N)=0~~~~for~~~ N<3
\end{equation}
So that, the production ratio of baryon to meson is obtained from
eq.~(\ref{w2}) and (\ref{w3}) 
\begin{equation}
\label{w5}
{R_{B/M} =} {{(1-a)N -b} \over {3(aN+b)}}
\end{equation}
From above, we see that QCM treats meson and
baryon formation uniformly, and there is no extra $ad~hoc$ 
mechanism and free
parameters for the baryon production. Here the $B/M$ ratio is
completely determined at a certain $N$, unlike in LM that it 
is completely uncertain and has to be adjusted by a free 
parameter~(the ratio of diquark to quark ${qq}/q$). 

\subsection {momentum distribution of primary 
hadrons in $q\overline{q}$ system}
\label{m2}

In order to give the momentum distribution of primary hadrons, each
phenomenological model must have some inputs. For example, in LM, 
they use a symmetric longitudinal fragmentation function
\begin{equation}
\label{lu}
f(z) \propto {{(1-z)^a} \over {z}} exp(-b{m_{T}^{2} \over z})
\end{equation}
where $a$ and $b$ are two free parameters (and $a$ is 
flavor dependent). In this paper, in order to give the momentum 
distribution of primary hadrons produced according to QPR and QCR, 
we simply adopt the widely used LPSA which is equivalent to the
constant distribution of rapidity. Hence a primary hadron $i$ is 
uniformly distributed in rapidity axis, then its rapidity can be 
written as
\begin{equation} 
\left \{
\begin{array}{l}
Y_{i}=Z+\xi_{i} Y \\
0 \leq \xi_{i} \leq 1
\end{array}
\right.
\end{equation} 
where $\xi_i$ is a random number; 
$Z$ and $Y$ are two arguments, and can be determined by
energy-momentum conservation in such color singlet system
\begin{equation}
\left \{
\begin{array}{l}
{\sum \limits_{i=1}^{H}} E_{i} = W \\
{\sum \limits_{i=1}^{H}} P_{Li} = 0
\end{array}
\right.
\end{equation} 
where $E_{i}$ and $P_{Li}$ denote the energy and the longitudinal 
momentum of the $i$th primary hadron respectively, they are 
obtained by
\begin {equation}
\left \{
\begin{array}{ll}
E_{i} = m _{Ti} {{exp(Y_{i}) + exp(-Y_{i})} \over {2}}\\
P_{Li}= m _{Ti} {{exp(Y_{i}) - exp(-Y_{i})} \over {2}}
\end{array}
\right.
\end{equation}
where $m_{Ti}$ is given by
\begin{equation} 
m _{Ti} = \sqrt{m_{i}^2 + {\stackrel {\rightarrow} P_{Ti}}^{2}}
\end{equation} 
where $m_{i}$ is the mass of the $i$th primary hadron, 
and $\stackrel {\rightarrow} P_{Ti}$ obeys the distribution
\begin{equation} 
\label{last}
f({\stackrel{\rightarrow} P_{T1}},\ldots,
{\stackrel{\rightarrow}P_{TH}}) \propto {\prod \limits_{i=1}^{H}}
exp(-{{{\stackrel {\rightarrow} P_{Ti}}^{2}} \over {\sigma^{2}}})
\delta ({\sum \limits_{i=1}^{H}} {\stackrel {\rightarrow} P_{Ti}})
\end{equation} 
In this paper, we set $\sigma=0.2~GeV$. Eq.~(\ref{last}) is 
just what LM uses. 

Note that LPSA or the constant rapidity distribution is rather 
naive, but it
is convenient for us to study the correlations without introducing
many parameters which would make the situation more complicated. 

\subsection {hadronization of multi-parton state}
\label{m3}
At the end of parton showering, a final multi-parton state
will start to hadronize. To connect the final multi-parton state 
with QCM, we adopt a simple treatment assumed in WM, i.e. before 
hadronization, each gluon at last splits into a $q'\overline{q'}$ 
pair, the $q'$ and $\overline{q'}$ carry one half of the gluon 
momentum, and each of them form a color singlet  with their 
counterpart antiquary and quark in their neighborhood, respectively.
Now take the three parton state $q\overline{q}g$ as an example to 
illustrate the hadronization of multi-parton state. Denote the 
4-momenta for $q$, $\overline{q}$, $g$ as
\begin{equation}
\left \{
\begin{array}{ll}
P_{1} & = ( E_{q} , {\stackrel {\rightarrow} P_{q}} )\\
P_{2} & = ( E_{g} , {\stackrel {\rightarrow} P_{g}} )\\
P_{3}&=(E_{\overline{q}},{\stackrel{\rightarrow}P_{\overline{q}}})
\end{array}
\right.
\end{equation}
Before hadronization, the gluon splits into a $q'\overline{q'}$ 
pair and the  $q'$ and $\overline{q'}$ carry one half of the 
gluon momentum, and the $q\overline{q}g$ system forms two 
color singlet subsystems $q\overline{q'}$ and $q'\overline{q}$. 
The invariant masses of the subsystems are
\begin{equation}
\left \{
\begin{array}{lll}
W_{q{\overline{q'}}} =& \sqrt {( P_{1} + {{P_{2}} \over {2}})^{2}}=
&\sqrt{(E_{q}+{{E_{g}}\over{2}})^{2}-({\stackrel{\rightarrow}P_{q}}+
{{\stackrel {\rightarrow} P_{g}} \over {2}})^{2}}\\
W_{{q'}{\overline{q}}} =&\sqrt {( P_{3} + {{P_{2}} \over {2}})^{2}} =
&\sqrt{(E_{\overline{q}}+{{E_{g}}\over{2}})^{2}-
({\stackrel{\rightarrow}P_{\overline{q}}}+
{{\stackrel {\rightarrow} P_{g}} \over {2}})^{2}}
\end{array}
\right.
\end{equation}
As was commonly argued by Sj\"{o}strand and Khoze 
recently$^{\cite{khoze}}$, the confinement effects should lead to a 
subdivision of the full $q\overline{q}$ system into color singlet 
subsystems with screened interactions between
these subsystems $q\overline{q'}$ and $q'\overline{q}$. 
Hence QCM can be applied independently to each color singlet 
subsystem, i.e., we can apply the equations in the former
two subsections to each subsystem, and obtain the momentum 
distribution for the primary hadrons in their own center-of-mass 
system. Then after Lorentz transformation, the momentum 
distribution of the  primary hadrons in laboratory frame is
given. This treatment can be extended to a general 
multi-parton state.

Obviously, when the emitting gluon is soft or collinear with the 
direction of $q$ or $\overline{q}$, $q\overline{q}g$ cannot be 
distinguished from $q\overline{q}$ and $W_{q\overline{q}'}$ or 
$W_{q'\overline{q}}$ is too small for hadronization. To avoid 
these cases, a cut-off mass $M_{min}$ has to be introduced. Here 
$M_{min}$ is a free parameter in perturbative phase. 
Its value and energy dependence is theoretically uncertain. 
The physical assumption is that $M_{min}$ is 
independent of energy$^{\cite{Mattig,sjostrand}}$. 

\section{Results and Comparison with data}
\label{iii}

As was mentioned above, to compare the predictions of a hadronization
model with data appropriately, a $e^+e^-\rightarrow h's$ generator is 
needed. The predictions of CMCP quoted in ref.~\cite{TASSO} was not 
obtained by a complete generator, since at least, it did not include 
the parton shower process. So it is necessary to recompare the 
predictions of QCM and TASSO data for $p\overline{p}$ rapidity 
correlation. Recently, OPAL collaboration have compared the
observed $\Lambda\overline{\Lambda}$ rapidity correlation with 
predictions of different hadronization models including LM, UCLA 
and WM by running the corresponding generators. 
Note that both LM and UCLA are embedded in JETSET. In order to 
compare with LM, UCLA and other models conveniently, we also 
replace LM with QCM in JETSET 7.3.  In this section, using this
modified JETSET, we give our predictions of $B\bar{B}$ rapidity 
correlation and other related properties, and compare them with 
the corresponding data.
All of our results in this paper are obtained by adjusting
only three energy independent parameters, i.e., the $M_{min}$ in the
perturbative phase, the $\beta$ in QPR and a spin suppression factor 
$\delta$ for decuplet baryon to octet one.

\subsection{strange baryon yields, $B/M$ ratio and momentum
distribution}

By choosing $M_{min}=2.6~GeV$, $\beta=4.2~GeV^{-1}$ and 
$\delta ={{(3/2)}^+\over{(1/2)}^+} =0.2$, we can 
describe most of the hadron yields. The predictions for them
in this paper are similar to that in ref.~\cite{qun}, so we will not
discuss them in detail here. OPAL collaboration have studied the 
strange baryon production. They find that even by adjusting 
the parameters which are related to the baryon production, 
JETSET~(LM) and HERWIG(WM) cannot fit all the data
well$^{\cite{Opal}}$. See table~\ref{yields} for their results. 
We also list our predictions in the same table. One can see that 
our results are better than those of LM and WM. According to the 
OPAL studies, the baryon yields is not sensitive to the popcorn 
parameter, and it is only responsible for the $B\overline{B}$ 
rapidity correlations. 

In $e^+e^-$ annihilation, because final hadrons observed in
experiments come from primary ones which are all hadronization 
results, the multiplicity ratio of primary baryons to mesons, 
known as the $B/M$ ratio is essential for understanding the 
universality of the hadronization mechanism. In LM, the $B/M$ 
ratio is additionally adjusted by a free parameter ${qq}/q$.
while in QCM, as is clearly shown in eq.~(\ref{w5}), 
it is completely determined.
So the predictions of the $B/M$ ratios are challenging for QCM. 
In table~\ref{y}, our predictions and the available 
data$^{\cite{group}}$ at $\sqrt{s} \simeq 30,~91~GeV$ are listed. 
It shows the agreement is surprisingly good.

Before studying the $B\overline{B}$ rapidity correlation, we 
should study the momentum distribution for the corresponding baryon. 
Since almost 70\% of TASSO data for 
$p\overline{p}$ correlations come from 34~GeV, 
we only give the differential cross section 
${d\sigma} \over {dP}$ of $p(\overline{p})$ at this energy~(see
fig.1a). To our surprise, with such a simple LPSA input,
the prediction of QCM for the momentum distribution agrees
with data in the region $P\geq 1.5$~GeV/c. But at 91~GeV, the
observed differential cross section of $\Lambda$ is found to be 
softer than any predictions of the JETSET~(LM), HERWIG~(WM) and QCM.
Particularly our prediction is even harder than LM~(fig.1b).
It certainly shows that the LPSA what we used is too naive to 
simulate the longitudinal distribution thoroughly.

\subsection {$B\overline{B}$ rapidity correlation}

The contradiction between TASSO data for $p\overline{p}$ 
rapidity correlation and the prediction of CMCP is always 
regarded as an evidence to rule out QCM.
In ref.~\cite{Xie1}, it was shown that QCM could be qualitatively 
consistent with data. But to compare with data quantitatively, the 
experimental conditions and multi-parton states must be taken into 
account. In this paper, we use the modified JETSET in which QCM is 
embedded to restudy the $p\overline{p}$ phase space correlation 
under the same conditions as in the
TASSO experiment~(the measurement is made for the proton with  
momenta between 0.4 and 1.2~GeV/c). 
Our results are shown in fig.2a,b. One can see that our predictions 
do agree with data. The predictions of LM, Meyer Model~(MM) and CMCP 
are also shown in fig.2a,b. Both LM and MM agree with data, too, 
while CMCP contradicts with data sharply. Note that our predictions 
seem to be the best  for such low statistics data.

Because of the limited statistics for the $p\overline{p}$ 
correlation, TASSO data cannot provide a high discriminating power 
among different models. Fortunately, OPAL collaboration recently 
present the high statistics data for the 
${\Lambda}\overline{\Lambda}$ rapidity correlation at 
91~GeV$^{\cite{Opalbb}}$. They compare their data with the 
predictions of LM, WM and UCLA. They find that the observed Rapidity 
Correlation Strength~(RCS)\footnote{The probability that 
$\overline{\Lambda}$ is found in an interval of $\pm 0.6$ around 
$\Lambda$ rapidity if the baryon number of a $\Lambda$ is compensated 
by a $\overline{\Lambda}$.} is weaker than that 
predicted by HERWIG and JETSET(LM) and stronger than that by the UCLA
model. The $\Lambda\overline{\Lambda}$ rapidity correlation can only 
be described by JETSET(LM) with the popcorn parameter 
($\rho={{BM\overline{B}}\over{B\overline{B}+BM\overline{B}}}$) 
adjusted to a rather high probability~(95\% popcorn)(fig.3a). 
This indicates that baryons appear from the successive production 
of several quark pairs in the popcorn scenario rather than that only
$B\overline{B}$ can emerge in the pure diquark model. Since part of 
time, the end result of the popcorn scenario will be exactly the
$B\overline{B}$  situation; however, further possibilities of the 
type $BM\overline{B}$, $BMM\overline{B}$, etc. are possible by color
fluctuation, i.e., a number of mesons are produced between $B$  and 
$\overline{B}^{\cite{sjostrand}}$. These configurations can be 
described naturally in QCM. In this sense, QCM is close to the 
popcorn scenario, and it should obtain the similar $B\overline{B}$ 
rapidity correlation. Our study does show this. In fig.3b, 
Our prediction for the $\Lambda\overline{\Lambda}$ 
rapidity correlation is given. The corresponding RCS is 53.4\%, 
which is in agreement with the OPAL measurement $\sim(53\pm 3)\%$.
Hence, without the popcorn and the diquark mechanism and any specific 
parameters, QCM does predict the $\Lambda\overline{\Lambda}$ rapidity 
correlation well. 

We also find that $B\overline{B}$ rapidity correlation is not 
sensitive to the parameters $M_{min}$ and $\beta$. First, we fix
$\beta=3.6~GeV^{-1}$~(the same value as in ref.~\cite{Xie2}) 
and change $M_{min}$ in a reasonable range; 
then we fix $M_{min}=2.6~GeV$ and change $\beta$. The predictions 
for $n_{\Lambda}$ and RCS at 91 GeV are listed in table~\ref{rcs1}.
We find that our predictions for $\Lambda\overline{\Lambda}$ rapidity
correlation is consistent with data even if the hadron 
multiplicities vary. This support a common belief that the 
$B\overline{B}$ rapidity correlation is more sensitive to the 
underlying fragmentation mechanisms than the inclusive properties.

\subsection {Local Baryon Number Compensation}

The PETRA and PEP experiments at $\sqrt{s} \simeq 30~GeV$ could 
demonstrate that the baryon number is dominantly conserved within 
the same hemisphere$^{\cite{petra,tassob}}$, which is called 
the local baryon number compensation. It is regarded as one 
consequence of chain-like models. But we find that it can also be 
explained by QCM naturally. In table~\ref{table}, we list our 
predictions of $pp$, $\overline{p}\overline{p}$ and 
$p\overline{p}$ pairs for proton~(or anti-proton) momenta 
between 1 and 5~GeV/c in the Same(S) or Different hemisphere(D) 
at $\sqrt{s}\simeq 30$GeV together with TASSO
data$^{\cite{tassob}}$. Our predictions are normalized to TASSO 
data. One can see that they are in agreement with  data. 
At $\sqrt{s}=91~GeV$, we calculate the ratio 
$R_{i}^{j}={{N_{i}^{j}} \over {\sum \limits_{k,l}}{N_{k}^{l}}}$ 
(for j,l=S, D and
i,k=$\Lambda\Lambda$($\overline{\Lambda}\overline{\Lambda}$),
$\Lambda\overline{\Lambda}$). Our predictions for 
$R_{\Lambda\Lambda(\overline{\Lambda}\overline{\Lambda})}^S$,
$R_{\Lambda\Lambda(\overline{\Lambda}\overline{\Lambda})}^D$, 
$R_{\Lambda\overline{\Lambda}}^S$ and 
$R_{\Lambda\overline{\Lambda}}^D$ 
are 10.1\%, 15.6\%, 61.1\% and 13.2\%, respectively. 
They indicate that the baryon
number tends to be conserved within the same hemisphere. 
These results can easily be understood from QCM. Before the 
hadronization of a multi-parton state,
a number of color singlet subsystems emerge. Each of them hadronizes
independently. According to the quark random combination picture, 
in the center-of-mass frame of each subsystem, baryons and 
anti-baryons must be produced in pairs and are distributed randomly 
in this subsystem, so the baryon number is compensated globally in 
it. This is just the case obtained by CMCP$^{\cite{Mattig}}$, since 
it only considered the hadronization of $q\overline{q}$ state. But 
each subsystem is one part of the whole system for the multi-parton 
state. In its center-of-mass frame, the baryon and anti-baryon 
produced in the same subsystem are generally much closer in phase 
space than those in different subsystems. 
Thus some local baryon number compensation can be obtained.

\section{Summary}
\label{iv}

In this paper, we use the PS program in JETSET to describe the parton 
shower process and to obtain the final multi-parton state via 
adjusting $M_{min}$. Before hadronization, the multi-parton state 
splits into a number of color singlet subsystems. In each subsystem,
we use QPR and QCR to describe the hadronization process, and give 
the momentum distribution of primary hadrons under LPSA. The unstable
hadron decays are handled by the corresponding simulation in JETSET. 
Since we only modify the program which describes the hadronization
phase, the global properties determined by other phases are the same 
as that given by the original JETSET 7.3. 

In describing meson production, QCM and LM are at the same level in 
respect of reproducing data and the number of adjustable parameters. 
But as was mentioned above, for the popular models, the trouble is to 
describe the baryon production, an aspect which more directly reflects 
the hadronization mechanism. They have to introduce additional 
mechanisms and corresponding parameters to describe the properties 
related to the baryon production. For instance, in LM, the $B/M$ ratio
depends on the diquark production and the free parameter ${qq}/q$ 
involved therein. The $B\overline{B}$ rapidity correlation is 
described by introducing the popcorn scenario and adjusting the free 
parameter $\rho$. The local baryon number compensation depends on the
chain-like picture. The strange baryon yields should be adjusted by 
other additional parameters, even so, the octet and decuplet cannot 
agree with data simultaneously.  In UCLA model, hadron yields can be 
reproduced well mainly by 5 parameters, but it cannot reproduce the
$B\overline{B}$ rapidity correlation well. HERWIG fails at both 
aspects though it reproduces global properties as well as JETSET.

Though QCM can describe baryon production naturally, it is still a 
critical problem whether QCM can reproduce the $B\overline{B}$ short
range correlation. In section~\ref{iii}, our results show QCM not 
only explain the $B\overline{B}$ short range rapidity correlation, 
but also give other properties that agree with data. Here, the 
$B\overline{B}$ rapidity correlation, the $B/M$ ratio and the local 
baryon number compensation are directly determined by the 
multi-parton state rather than by additional specific mechanisms and
associated free parameters. In QCM, these properties are related to 
the average number $<N>$ of quark pairs in each color 
singlet subsystem which is controlled by $M_{min}$ and $\beta$. 
When $<N>$ decreases, the $B/M$ ratio will be smaller,  
the RCS become stronger, and the local baryon number compensation 
appears more obvious. This reveals that these phenomena 
that seem to have no relations result from a common origin which is 
directly connected with the multi-parton state before hadronization.
Therefore we conclude that QCM not only cannot be ruled out by 
$B\overline{B}$ correlations, but also provides a good 
understanding for them. This indicates that QCM seems to be  
a very promising picture for hadronization.

\begin{center} \bf{ Figure Captions} \end{center}

\bf{Fig.1} \bf{(a)} Differential cross section as a function of 
hadron momenta(P) for $p$($\overline{p}$) at 34 GeV. TASSO data 
and the prediction of QCM are shown. \bf{(b)} Differential cross 
section as a function of $X_{E} = {{2E_{hadron}} \over {\sqrt{s}}}$ 
for $\Lambda$ at 91 GeV.
The curves show the respective Monte Carlo differential cross 
sections from LM$^{\cite{Opal}}$ and QCM.

\bf{Fig.2} The $p\overline{p}$ correlations for \bf{(a)} 
difference of $p$ and
$\overline{p}$ in rapidity, \bf{(b)} $\cos\theta$, where
$\theta$ is the angle between $p$ and $\overline{p}$. 
The predictions of QCM, LM, CMCP and MM is also drawn 
together with TASSO data$^{\cite{TASSO}}$.

\bf{Fig.3}: Rapidity difference for all $\Lambda\overline{\Lambda}$ 
pairs at 91 GeV. The distributions expected from \bf{(a)} LM with 
no popcorn, LM with 95\% popcorn, UCLA model and 
WM$^{\cite{Opalbb}}$, \bf{(b)} QCM.


\begin{table}[t]
\caption{Inclusive strange hadron yields by OPAL and the predictions 
from LM, WM$^{\cite{Opal}}$ and QCM.}
\begin{tabular}{|c|c|c|c|c|c|c|}
Particle & ${n}\over{event}$(OPAL) & LM default & LM tuned 
& WM default &WM tuned & QCM \\ \hline
$\Lambda$ & $0.351\pm 0.019$ &0.383 & 0.351 & 0.427 & 0.352&0.385\\
\hline
$\Xi$ & $0.0206\pm0.0021$ & 0.027& 0.021& 0.062&0.046&0.042\\ \hline
$\Sigma(1385)^\pm$&$0.0380\pm 0.0062$&0.074&0.068&0.136&0.115&0.0389
\\ \hline
$\Xi(1530)^0$ &$0.0063\pm 0.0014$ &0.0053 &0.0048 &0.0307 &0.0216 
&0.0064 \\ \hline
$\Omega$ &$0.0050\pm0.0015$ &0.00072 &0.00044&0.0095&0.0054&0.0043\\
\end{tabular}
\label{yields}
\end{table}

\begin{table}[t]
\caption{The $B/M$ ratio  at 
$\sqrt{s} \simeq 30~GeV$ and $\sqrt{s}=91~GeV$.}
\begin{tabular}{|c|cc|cc|} 
  &$\sqrt{s}\simeq$ 30~GeV &  &$\sqrt{s}$=91~GeV &  \\ 
 & QCM &data$^{\cite{group}}$ &QCM &data$^{\cite{group}}$ \\ \hline
$p \over {\pi^{+}}$  &0.058 &$0.062 \pm 0.005$ & 0.051 
 &$0.054 \pm 0.007$\\ \hline
$P \over {K^{+}}$   & 0.457 &$0.432\pm 0.043$ & 0.45 
    &$0.38 \pm 0.05$  \\ \hline
$\Lambda \over {{K}^{+}}$  &0.201 &$0.139 \pm 0.011$  
        & 0.203 & $0.144 \pm 0.005$ \\ \hline
$\Sigma(1385)^\pm \over K^{\star +}$  &0.0442 & $0.0516\pm 0.0131$  
&0.0421 &$0.0487\pm 0.0094$ \\ \hline
$\Xi(1530)^0 \over K^{\star 0}$  &0.0075 &  
&0.007 &$0.0082\pm 0.0021$  \\ 
\end{tabular}
\label{y}
\end{table}

\begin{table}[t]
\caption{Our predictions of $n_{\Lambda}$ and RCS 
for different $M_{min}$ at fixed $\beta=3.6$~GeV$^{-1}$ and
for different $\beta$ at fixed $M_{min}=2.6$~GeV.}
\begin{tabular}{|ccccc|ccccc|}
  &  & $\beta=3.6$ GeV$^{-1}$& & & & $M_{min}=2.6$ GeV & & &\\ \hline
$M_{min}$(GeV)  &  2     &  2.6   & 3     & 4  
& $\beta(GeV^{-1})$ & 3.0 & 3.6 & 4.2 & 4.8 \\ \hline
$n_{\Lambda}$ & 0.289 & 0.362  & 0.399 & 0.492 
& &0.337 & 0.362 & 0.385 &0.409\\ \hline
RCS(\%)      & 54.6   & 52.1   & 51.6  & 48.4 
 &  & 53.2 &52.1 &53.4 &51.9\\ 
\end{tabular}
\label{rcs1}
\end{table}

\begin{table}[t]
\caption{Our predictions for $pp$, $\overline{p}\overline{p}$ 
and $p\overline{p}$ in the same~(S) or
different~(D) hemisphere at~$\sim$30~GeV together with TASSO data}
\begin{tabular}{|c|cc|cc|} 
        & S    &      &  D  & \\ \hline
        & QCM  & data & QCM &data \\ \hline
$pp(\overline{p}\overline{p})$ & 1.8&$1.5\pm2.1$
& 3.5&$3.5\pm2.9$ \\ \hline
$p\overline{p}$        & 12.8&$15.5\pm4.5$ &3.6&$1.2\pm2.6$ \\ 
\end{tabular}
\label{table}
\end{table}

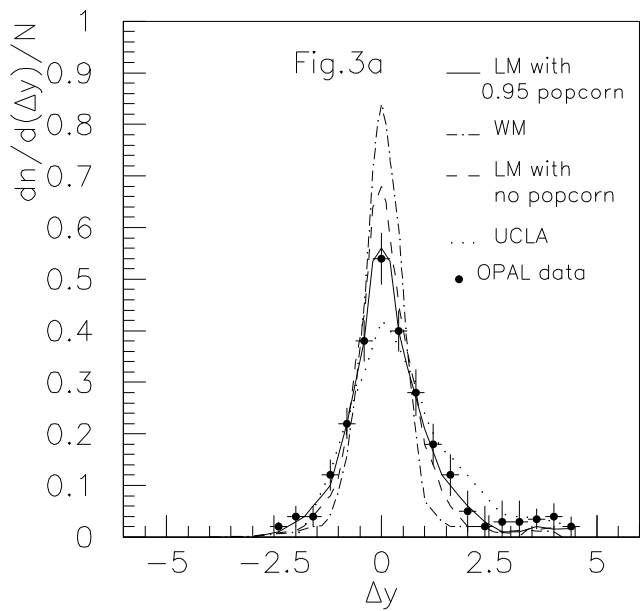
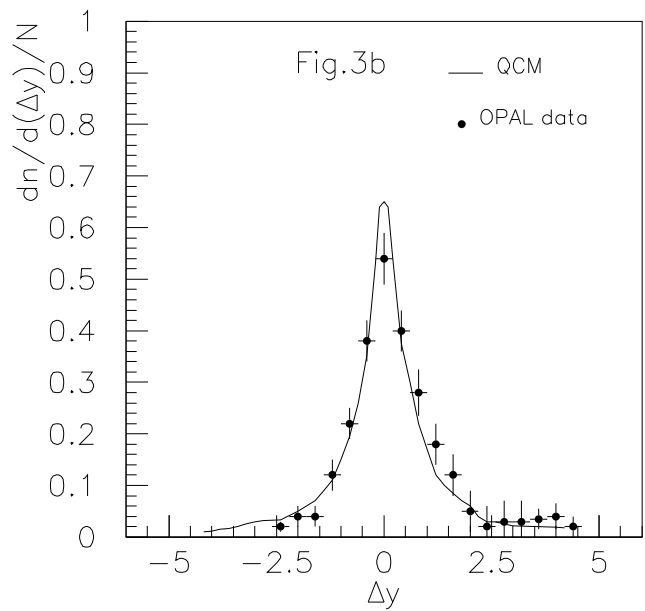
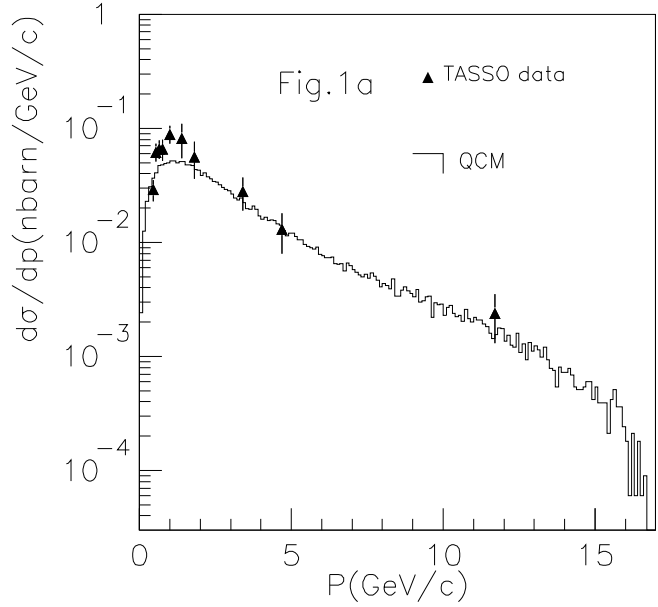
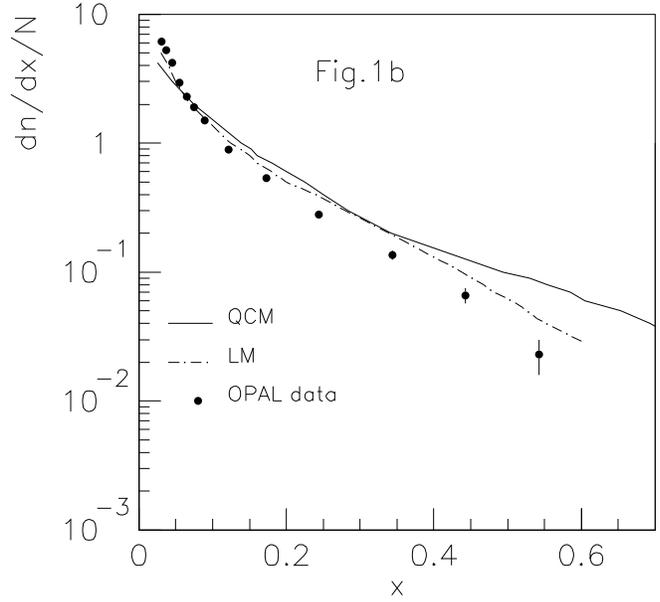
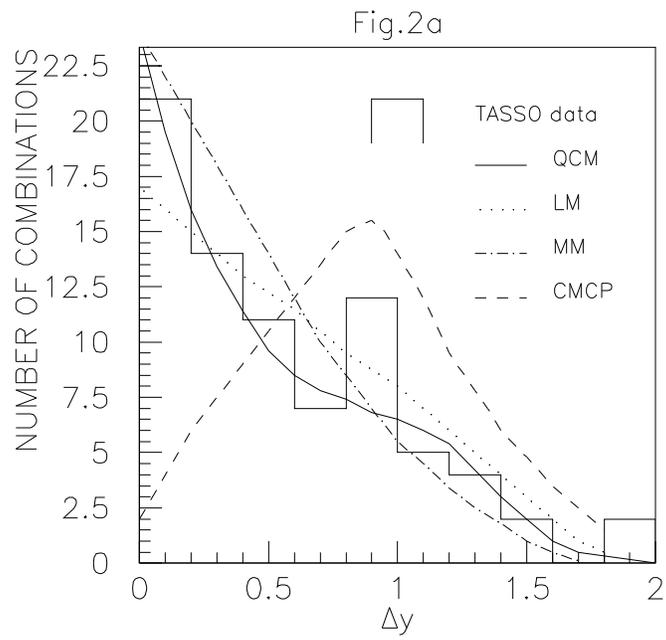
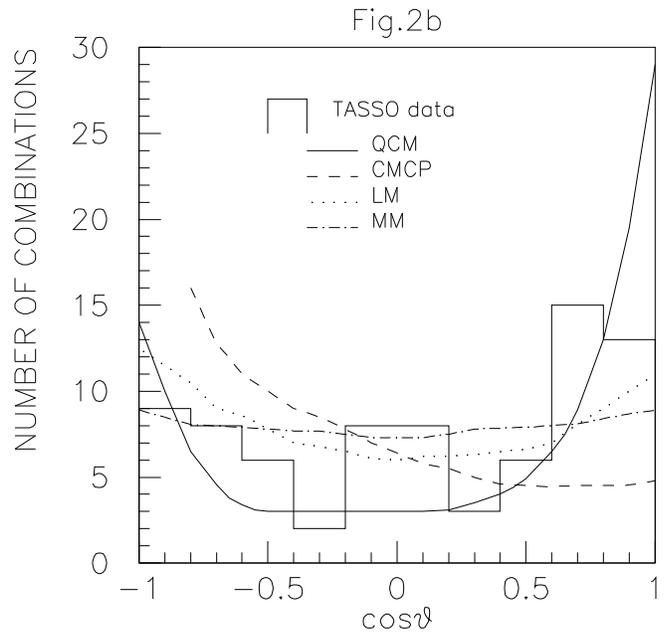
\end{document}